\documentclass[conference]{IEEEtran}      

\usepackage[bookmarks=false]{hyperref} 
\hypersetup{draft}
\usepackage[utf8]{inputenc}
\usepackage{amsmath,amsfonts,amsbsy,amssymb}
\usepackage{mathabx}
\usepackage{mathrsfs}
\usepackage{tabularx}
\usepackage{graphicx}
\usepackage{cite}
\usepackage{wasysym}
\usepackage{multirow}
\usepackage{float}
\usepackage{color}
\usepackage{epstopdf}
\usepackage{mathtools}
\usepackage{graphics} 
\usepackage{graphicx}
\usepackage{siunitx}
\usepackage[nolist]{acronym}
\usepackage{siunitx}
\usepackage{eurosym}
\usepackage{footnote}

\definecolor{green1}{RGB}{70, 180, 80}

\title{Brain-to-Brain Communication Based on Wireless Technologies: Actual and Future Perspectives}

\author{Dick Carrillo$^{1}$, Renan Moioli$^{2}$, Pedro Nardelli$^{1}$\\
$^{1}$ School of Energy Systems, LUT University, Finland\\
$^{2}$ Digital Metropolis Institute, Federal University of Rio Grande do Norte, Brazil\\
}

\begin{document}
\maketitle
\thispagestyle{empty}
\pagestyle{empty}

\begin{abstract}
During the last few years, intensive research efforts are being done in the field of brain interfaces to extract neuro-information from the signals representing neuronal activities in the human brain. 
Recent development of brain-to-computer interfaces support direct communication between animals' brains, enabling direct brain-to-brain communication.
Although these results are based on binary communication with relaxed requirements of latency and throughput, the fast development in neuro-science technologies indicates potential new scenarios for wireless communications between brains.

In this paper we highlight technologies that are being used today to enable brain-to-brain communication and propose potential wireless communication architectures and requirements for future scenarios.

\end{abstract}
\begin{acronym}
  \acro{1G}{first generation of mobile network}
  \acro{1PPS}{1 pulse per second}
  \acro{2G}{second generation of mobile network}
  \acro{3G}{third generation of mobile network}
  \acro{4G}{fourth generation of mobile network}
  \acro{5G}{fifth generation of mobile network}
  \acro{ARQ}{Automatic repeat request}
  \acro{ASIP}{Application Specific Integrated Processors}
  \acro{AWGN}{additive white Gaussian noise}
   \acro{BER}{bit error rate}
  \acro{BCH}{Bose-Chaudhuri-Hocquenghem}
  \acro{BRIC}{Brazil-Russia-India-China}
  \acro{BS}{base station}
  \acro{CDF}{Cumulative Density Function}
  \acro{CoMP} {cooperative multi-point}
  \acro{CP}{cyclic prefix}
  \acro{CR}{cognitive radio}
  \acro{CS}{cyclic suffix}
  \acro{CSI}{channel state information}
  \acro{CSMA}{carrier sense multiple access}
  \acro{DFT}{discrete Fourier transform}
  \acro{DFT-s-OFDM}{DFT spread OFDM}
  \acro{DSA}{dynamic spectrum access}
  \acro{DVB}{digital video broadcast}
  \acro{DZT}{discrete Zak transform}
  \acro{eMBB} {Enhanced Mobile Broadband}
  \acro{EPC}{evolved packet core}
  \acro{FBMC}{filterbank multicarrier}
  \acro{FDE}{frequency-domain equalization}
  \acro{FDMA}{frequency division multiple access}
  \acro{FD-OQAM-GFDM}{frequency-domain OQAM-GFDM}
  \acro{FEC}{forward error control}
  \acro{F-OFDM}{Filtered Orthogonal Frequency Division Multiplexing}
  \acro{FPGA}{Field Programmable Gate Array}
  \acro{FTN}{Faster than Nyquist}
  \acro{FT}{Fourier transform}
  \acro{FSC}{frequency-selective channel}
  \acro{GFDM}{Generalized Frequency Division Multiplexing}
  \acro{GPS}{global positioning system}
  \acro{GS-GFDM}{guard-symbol GFDM}
  \acro{IARA}{Internet Access for Remote Areas}
  \acro{ICI}{intercarrier interference}
  \acro{IDFT}{Inverse Discrete Fourier Transform}
  \acro{IFI}{inter-frame interference}
  \acro{IMS}{IP multimedia subsystem}
  \acro{IoT}{Internet of Things}
  \acro{IP}{Internet Protocol}
  \acro{ISI}{intersymbol interference}
  \acro{IUI}{inter-user interference}
  \acro{LDPC}{low-density parity check}
  \acro{LLR}{log-likelihood ratio}
  \acro{LMMSE}{linear minimum mean square error}
  \acro{LTE}{long-term evolution}
  \acro{LTE-A}{Long-Term Evolution - Advanced}
  \acro{M2M}{Machine-to-Machine}
  \acro{MA}{multiple access}
  \acro{MAR}{mobile autonomous reporting}
  \acro{MF}{Matched filter}
  \acro{MIMO}{multiple-input multiple-output}
  \acro{MMSE}{minimum mean square error}
  \acro{MRC}{maximum ratio combiner}
  \acro{MSE}{mean-squared error}
  \acro{MTC}{Machine-Type Communication}
  \acro{NEF}{noise enhancement factor}
  \acro{NFV}{network functions virtualization}
  \acro{OFDM}{Orthogonal Frequency Division Multiplexing}
  \acro{OOB}{out-of-band}
  \acro{OOBE}{out-of-band emission}
  \acro{OQAM}{Offset Quadrature Amplitude Modulation}
  \acro{PAPR}{Peak to average power ratio}
  \acro{PDF}{probability density function}
  \acro{PHY}{physical layer}
  \acro{QAM}{quadrature amplitude modulation}
  \acro{PSD}{power spectrum density}
  \acro{QoE}{quality of experience}
  \acro{QoS}{quality of service}
  \acro{RC}{raised cosine}
  \acro{RRC}{root raised cosine}
  \acro{RTT} {round trip time}  
  \acro{SC}{single carrier}
  \acro{SC-FDE}{Single Carrier Frequency Domain Equalization}
  \acro{SC-FDMA}{Single Carrier Frequency Domain Multiple Access}
  \acro{SDN}{software-defined network}
  \acro{SDR}{software-defined radio}
  \acro{SDW}{software-defined waveform}
  \acro{SEP}{symbol error probability}
  \acro{SER}{symbol error rate}
  \acro{SIC}{successive interference cancellation}
  \acro{SINR}{signal-to-interference-and-noise ratio }
  \acro{SMS}{Short Message Service}
  \acro{SNR}{signal-to-noise ratio}
  \acro{STC}{space time code}
  \acro{STFT}{short-time Fourier transform}
  \acro{TD-OQAM-GFDM}{time-domain OQAM-GFDM}
  \acro{TTI}{time transmission interval}
  \acro{TR-STC}{Time-Reverse Space Time Coding}
  \acro{TR-STC-GFDMA}{TR-STC Generalized Frequency Division Multiple Access}
  \acro{TVC}{ime-variant channel}
  \acro{UFMC}{universal filtered multi-carrier}
  \acro{UF-OFDM}{Universal Filtered Orthogonal Frequency Multiplexing}
  \acro{UHF}{ultra high frequency}
  \acro{URLL}{Ultra Reliable Low Latency}
  \acro{V2V}{vehicle-to-vehicle}
  \acro{V-OFDM}{Vector OFDM}
  \acro{ZF}{zero-forcing}
  \acro{ZMCSC}{zero-mean circular symmetric complex Gaussian}
  \acro{W-GFDM}{windowed GFDM}
  \acro{WHT}{Walsh-Hadamard Transform}
  \acro{WLAN}{wireless Local Area Network}
  \acro{WLE}{widely linear equalizer}
  \acro{WLP}{wide linear processing}
  \acro{WRAN}{Wireless Regional Area Network}
  \acro{WSN}{wireless sensor networks}
  \acro{ROI}{return on investment}
  \acro{NR}{new radio}
  \acro{SAE}{system architecture evolution}
  \acro{E-UTRAN}{evolved UTRAN}
  \acro{3GPP}{3rd Generation Partnership Project }
  \acro{MME}{mobility management entity}
  \acro{S-GW}{serving gateway}
  \acro{P-GW}{packet-data network gateway}
  \acro{eNodeB}{evolved NodeB}
  \acro{UE}{user equipment}
  \acro{DL}{downlink}
  \acro{UL}{uplink}
  \acro{LSM}{link-to-system mapping}
  \acro{PDSCH}{physical downlink shared channel}
  \acro{TB}{transport block}
  \acro{MCS}{modulation code scheme}
  \acro{ECR}{effective code rate}
  \acro{BLER}{block error rate}
  \acro{CCI}{co-channel interference}
  \acro{OFDMA}{orthogonal frequency-division multiple access}
  \acro{LOS}{line-of-sight}
  \acro{VHF}{very high frequency}
  \acro{pdf}{probability density function}
  \acro{ns-3}{Network simulator 3}
  \acro{Mbps}{mega bits per second}
  \acro{IPC}{industrial personal computer}
  \acro{RSSI}{received signal strength indicator}
  \acro{OPEX}{operational expenditures}
  \acro{H2H}{human-to-human}
  \acro{DOD}{depth of discharge}
  \acro{ADC}{analog-to-digital converter}
  \acro{FFT}{Fourier fast transform}
  \acro{DAC}{digital-to-analog converter}
  \acro{IFFT}{inverse Fourier fast transform}
  \acro{DC/DC}{Direct-to-Direct current converter}
  \acro{BBI}{brain-to-brain interface}
  \acro{B2BC}{brain-to-brain communication}
  \acro{SON}{self-organized networks}
  \acro{NOMA}{non-orthogonal multiple access}
  \acro{UART}{universal asynchronous receiver transmitter}
\end{acronym}

\renewcommand\IEEEkeywordsname{Keywords}
\section{Introduction}
In \ac{B2BC} the information is obtained by interfaces that combine neurostimulation, neuroimaging, and neuromodulation methods to obtain and deliver information between brains, allowing direct brain-to-brain communication. 
In very basic terms, information is extracted from the neural signals of a sender brain, digitized, and then delivered  to a receiver brain. Importantly, electrical or behavioral feedback improve performance of both subjects.
Early  interest  in \ac{B2BC} came  from  the  potential for expanding  human  communication  and  social interaction capabilities \cite{ref_brain_1, ref_brain_2,pais2013brain,pais2015building}.
However, \ac{B2BC} is still in its infancy and lack several key features of real-world human communication. For example, subject interactivity has been minimal.

On a parallel front, new advances in \ac{B2BC} have been designed to support communication for more than two human subjects. 
One possible scenario is composed by a communication structure that supports two Senders and one Receiver.
However, based on authors statement, it can be readily scaled up to include multiple Senders.
%
%
Additionally, researchers are currently exploring the use of novel techniques as functional magnetic resonance imaging (fMRI) that will increase the bandwidth of data transmitted over brain-to-computer interfaces (BCI) \cite{ref_b2b_com_1,ref_b2b_com_2}. 
%

Nevertheless, the recent \ac{B2BC} results are still characterized by restricted binary information with extreme low data rates without any latency requirements.
In this sense, we would like here to point out potential requirements for future wireless communication scenarios \cite{ref_wireless_b2b} where different nodes (Each node represents a \ac{B2BC} user) inside a specific wireless coverage are interchanging reliable information through a variety of communication mechanisms (either half-duplex or full-duplex communication).
%
%
In this context, new challenges for wireless technologies need to be addressed.
Specially, at the initial stages, an important requirement will be related to data security, latency, and reliability.
Other further steps will be related to network issues, for example, users density as an additional requirement to support higher data rates and low data rates.

It is important to remark that the aim of this paper is not related to signal processing techniques widely used in brain-to-computer interfaces.
Our focus is associated to the communication network used to transmit the neuro-information.
%

%
This paper is organized in following sections.
In Section II, we provide a summary of actual architectures and wireless technologies used today in \ac{B2BC}.
Future perspectives for new Architectures and requirements for \ac{B2BC} are described in Section III.
Finally, a brief summary is done in Section IV.

\section{Wireless Communication Actuality in \ac{B2BC}}
Nowadays, most effort in \ac{B2BC} is concentrated in the signal processing and techniques that improve brain signals interpretation.
So, communication technologies are not a key enabler for actual \ac{B2BC} development because data transmitted is used with relaxed requirements and most of the cases the communication is reduced to transmit e-mail messages through Internet.
However, it is important for practitioners and engineers to get information of actual technologies that support the present development of \ac{B2BC}.
This architecture and technologies are cited in following lines.
\subsection{Architecture Used Today in \ac{B2BC}}
Most of actual architectures, from the communication point of view, are limited to one way communication \cite{ref_b2b_com_1,ref_b2b_com_2}.
It means that the data information goes only on one way from transmitter and receiver without any feedback or answer of the receiver.
For example, in \cite{ref_b2b_com_2} the architecture is composed basically by the following architecture elements: emitter, computer, communication channel, receiver.
%
\subsubsection{Emitter}
The emitter is the one that generates the information extracted from the brain, this information is transmitted to the computer using a wireless serial communication as Bluetooth or in some cases it could be wired \ac{UART} communication.

\subsubsection{Computer}
The computer is used to process the brain-to-computer and computer-to-brain communication in the emitter and receiver side respectively. 

\subsubsection{Communication Channel}
The information generated by computer is transmitted through Internet to other computer element using any commercial communication technology (wired or wireless) as ethernet, wifi, \ac{LTE}.
All these cases without any important requirements because most of the data transmitted are reduced to small packets without any latency and throughput requirement.

\subsubsection{Receiver}
In the other side, the information receiver subjects were stimulated with biphasic transcranial magnetic stimulation pulses at a subject-specific occipital cortex site.
The intensity of pulses was adjusted for each subject so that a) one particular orientation of the transcranial magnetic stimulation-induced electric field produced phosphenes \cite{ref_b2b_com_2} (representing the active direction and coding the bit value ‘‘1’’), and b) the orthogonal direction did not produce phosphenes (representing the ‘‘silent direction’’ and coding the bit value ‘‘0’’). Subjects reported verbally whether or not they perceived phosphenes on stimulation.

\begin{figure}[ht]
   \centering
   \includegraphics[width=\columnwidth]{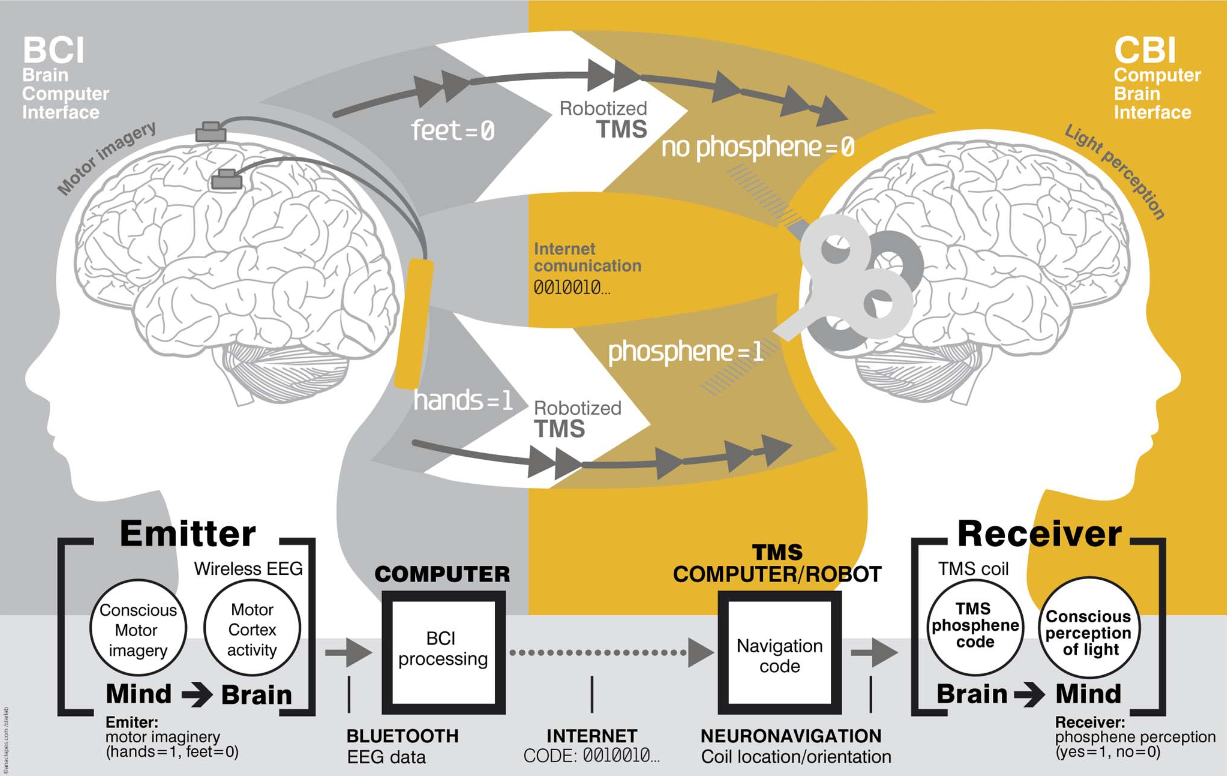}
   \caption{Brain-to-brain communication architecture proposed in \cite{ref_b2b_com_2}. 
   }
   \label{fig:b2b_presentt}
\end{figure}

A diagram published in \cite{ref_b2b_com_2} is shown in Figure \ref{fig:b2b_presentt}. 
Here the brain-to-computer interface subsystem is shown schematically, including electrodes over the motor cortex and the electroencephalographic amplifier/transmitter wireless box in the cap. Motor imagery of the feet codes the bit value 0, of the hands codes bit value 1. On the right of the same figure, the computer-to-brain interface system is illustrated, highlighting the role of coil orientation for encoding the two bit values. Communication between the  brain-to-computer interface and computer-to-brain interface components is mediated by the internet.

As it was explained before, the actual architecture is simplified from the communication network point of view.
In which most of the effort is concentrated in the signal processing developed to support the emitter and the receiver sides.
\subsection{Wireless Technologies used in Actuality}
From the communication network perspective, ethernet is the most important technology that is being used in actual projects.
this protocol is typically running on wired connectivity and in some cases using wireless.
In general, as there is not specific requirements to transmit information from brain-to-brain interfaces, the choice depends of what technology is available.
Without paying attention to latency, reliability, throughput.
In some cases the QoS (quality of service) for these type of applications could be simplified to full buffer traffic.
It means, that data is transmitted over the communication network without any requirement, with the possibility to become the data packet with the worst priority.
For illustrative purposes, we list today's technologies with the most influence and their respective characteristics.
\begin{itemize}
\item WiFi, today the most popular technology that operates in unlicensed bands. The latest standard (IEEE 802.11) release considers \ac{OFDM} as modulation technology. The multiple access technology is based on Carrier-sense multiple access with collision avoidance (CSMA-CA), and  it also supports \ac{MIMO}. WiFi operates typically on interference conditions, which is a drawback in some applications \cite{wifibook}.
\item Bluetooth, other popular technology developed to support wireless serial communication in short distances. It is based on IEEE 802.15.1 standard. It could be considered to build personal area networks (PANs). It was originally conceived as a wireless alternative to RS-232 data cables. In the first releases it was based on Gaussian frequency-shift keying (GFSK) modulation. Since the introduction of Bluetooth 2.0+EDR, $\pi/4$-DQPSK (differential quadrature phase-shift keying) and 8-DPSK modulation may also be used between compatible devices. Devices functioning with GFSK are said to be operating in basic rate (BR) mode where an instantaneous bit rate of 1 Mbit/s is possible. The term Enhanced Data Rate (EDR) is used to describe $\pi/4$-DPSK and 8-DPSK schemes, each giving 2 and 3 Mbit/s respectively. The combination of these (BR and EDR) modes in Bluetooth radio technology is classified as a BR/EDR radio \cite{bluetooth}. 
\item \ac{LTE}, the cellular technology that is widely installed around the world. It is based on the 3GPP Rel 8 and following releases. The waveform and multiplexing scheme is based on \ac{OFDMA}, it also support \ac{MIMO}. In contrast of other technologies, \ac{LTE} is a based on a centralized communication in the base station. This base station manage different network impairments as interference, adaptive modulation, power transmission management, resource allocation, and other features that makes \ac{LTE} a technology reliable that support different \ac{QoS} policies \cite{ltebook}.
\item proprietary RF, the previous technologies are based on any standard. But in some specific cases some technologies are developed to support specific requirements. In some cases they are based on specific modulation schemes that are adapted to specific scenarios. Companies developed their own technology to support specific applications, as for example \ac{B2BC}.
\end{itemize}
With the intention to contribute with practitioners and engineers, the Table \ref{table:today_technologies} provides a list of prof and cons of previous technologies explained before.
\begin{table}
\centering
\caption{Wireless Technologies showing pros and cons}
\label{table:today_technologies}
\begin{tabular}{|p{2cm}||p{2cm}||p{2cm}|}
 \hline
 \multicolumn{3}{|c|}{Pros and con of wireless technologies in \ac{B2BC}} \\
 \hline
 \hline
 \textbf{Technology}& \textbf{Pros} & \textbf{Cons}\\
 \hline
 \hline
 WiFi& Cheaper, Popular    &high interference\\
 \hline
 Bluetooth&   cheaper  & small distances   \\
 \hline
 LTE& world wide installed, good coverage & expensive\\
 \hline
 Proprietary RF& optimized for specific application& expensive\\
\hline 
\end{tabular}
\end{table}

\section{Future Wireless Technologies in \ac{B2BC}}
It is not difficult to realize the potential applications that \ac{B2BC} will introduce in the society.
For example, the rehabilitation of persons suffering from poor motor conditions is one of the core drivers that motivate research in this field.
It is easy to see that not health applications would benefit from the development of \ac{B2BC}  but also 
industrial application in robotics, automation, entertainment, and social relations may also become scenarios where \ac{B2BC} could be the next (revolutionary) step after human-computer interfaces based on virtual/augmented reality and social media interactions. 
In this context, in following lines we provide some insights of network architectures and requirements that potentially will support the future applications of \ac{B2BC}.
\subsection{Potential Network Architecture for \ac{B2BC}}
It is expected that future evolution of \ac{B2BC} will demand full-duplex and half-duplex communication, as a contrast of what is used today.
It also is expected that traditional cellular architectures could be useful in some scenarios (centralized communication in a specific base station).
In some scenarios, a mesh topology with few users could be a good option.
In all cases, it is expected that \ac{B2BC} will be realized mostly in high dense user per area coverage.
So, potentially mmWave combined with beamforming will be considered to support radio access connectivity with user-centric virtual cell in ultra dense networks \cite{ref_beamforming}.

In specific scenarios, for example medical applications, \ac{B2BC} should be done using uniquely one sender and one receiver.
So, in this ideal scenario, many network layers could be avoided in order to optimize latency and throughput.
For example, uplink/downlink resource allocation algorithms could be avoided.
Other feature that could be avoided is to up scale to \ac{IP} because Internet will not be an important requirement for these applications. 

\begin{figure*}[!t]
	\centering
	\includegraphics[width=2\columnwidth]{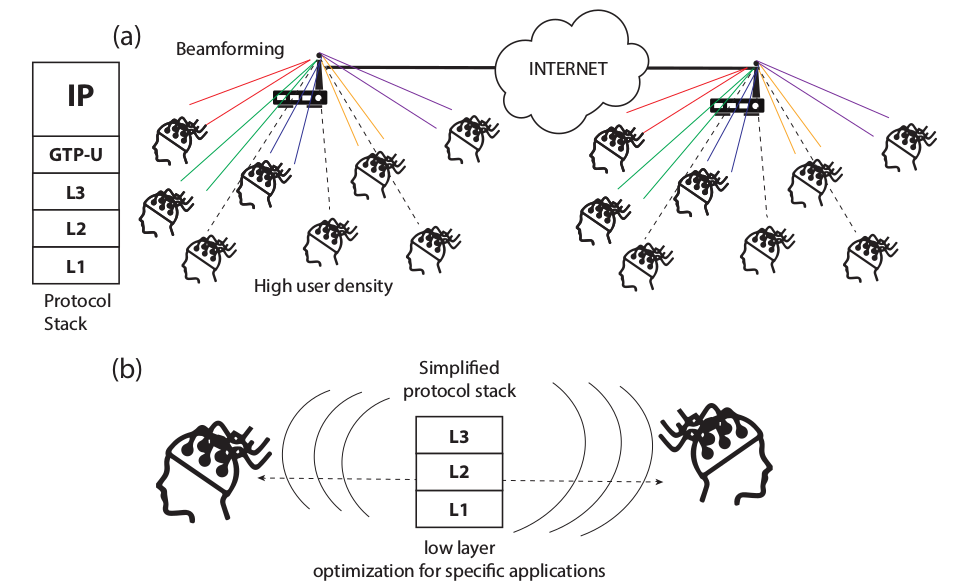}
	\caption{\label{fig:system1}(a) Architecture point-to-multipoint (cellular architecture) with Internet integration that is possible with the usability of multiple layer communications to support typical network impairments. (b) simplified protocol stack to point-to-point communication.}
	\vspace{-1ex}
\end{figure*}

Both previous mentioned architectures are described in the diagrams of Fig. \ref{fig:system1}.

Independently of network architecture, security issues should be the most important factor that will be considered in \ac{B2BC}.
This new security challenges aims to design efficient secure transmission schemes that exploit propagation properties of radio channels.
With careful management and implementation, physical layer security and conventional encryption techniques can formulate a well-integrated security solution together that efficiently safeguards the confidential and privacy communication data \cite{physecurity}.

Other important aspect of potential network architecture in \ac{B2BC} is the utilization of local area data network (LADN) that geographically isolates the operator network resources to provide high data rate, low latency,
and service localization for the wireless System Architecture \cite{LDAN}.

It is a fact that a variety of architectures should be supported in order to provide flexibility to the radio resource management (optimizing wireless channel resources).
Finally, it is expected that many wireless technologies will be capable to support \ac{B2BC}.
It means that co-existing scenarios should be considered, specially to optimize physical layer to full interference scenarios.
\begin{figure}[!t]
	\centering
	\includegraphics[width=.8\columnwidth]{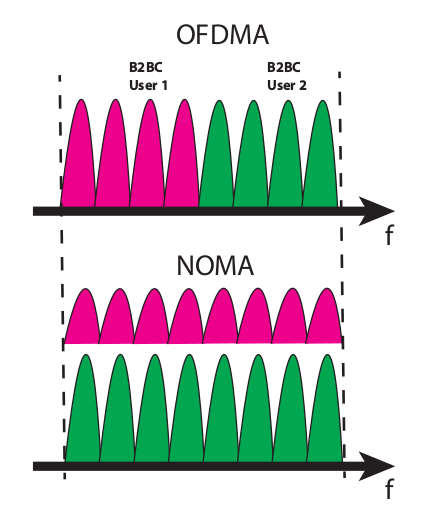}
	\caption{\label{fig:noma}Comparison of Multiple access technologies with special emphasis in \ac{NOMA} that is useful in scenarios with strong interference because high user density. In \ac{NOMA} users can use the same resource block simultaneously.}
\end{figure}

\subsection{Potential \ac{B2BC} Technology Enablers}
Considering that \ac{B2BC} will achieve complex scenarios with a diversity of network impairments.
We provide some insights of potentially technology enablers of the future \ac{B2BC} in following lines.
\subsubsection{\ac{GFDM}}
It is a flexible waveform that is a generalization of the popular \ac{OFDM} \cite{ref_GFDM_dick}. 
This flexibility enables the possibility to support different data rates and different latency values that is possible because \ac{GFDM} is a generalization of \ac{OFDM} with many optional configurations of subcarriers and subsymbols.
Other important advantage is that in the latest years many research investment was done in this waveform to improve \ac{GFDM} encoder performance.

\subsubsection{Non-Orthogonal Multiple Access}
Scenarios with high user density decrease the performance of the system.
However, when \ac{NOMA} is used, many research results have shown that the capacity network could be improved \cite{ref_noma_3} (See Fig. \ref{fig:noma}).
In the case of beyond \ac{5G}, we expect to apply hybrid solutions using \ac{NOMA} methodologies to support the various scenarios of \ac{B2BC}.

\subsubsection{Artificial Intelligence}
Digital signal processing is a key tool in most of wireless communication technologies.
However, artificial intelligence techniques are gaining popularity, even in detection, channel estimation, radio resource scheduling, and power optimization \cite{ref_intel_artif}.
So, it is expected that popularity of artificial intelligence will provide a positive impact of \ac{B2BC}, specially to optimize dynamically the network performance.
%
%
\subsubsection{\ac{URLL} in mmWaves}
The cellular generation \ac{5G} has two important key technologies to support independently industry requirements, they are \ac{URLL} and mmWaves (millimeter Waves).
However, to support \ac{B2BC} it will be necessary that both technologies will join features in order to enable high level quality of service requirements.
\subsubsection{Local Core Network}
With the improvement of processors and hardware, it will be possible in beyond \ac{5G} technologies to run local core networks without the necessity to operate with a backhaul connectivity, as is used today.
This feature, will have strong consequences in the business model and application of these future networks \cite{5G_model_iran}.
This independence does not imply that features related to network management will be neglected. 
However, this networks will gain the popularity of WiFi with the advantage to support advanced features.

\subsubsection{3D Beamforming with MIMO}
the beamforming of 3D-MIMO can be designed in full 3D space, which can substantially improve the system capacity and alleviate the multi-user interference.
The performance of beamforming relies heavily on precise CSI. However, in realistic scenarios, the channel knowledge is generally imperfect. 
Therefore, CSI uncertainty should be considered in beamforming design so that the system performance is robust to imperfect channels \cite{3dbeamforming}.

\vspace{-.2cm}
\section{Summary}
Nowadays, \ac{B2BC} is becoming an interesting research topic, with special attention to signal processing used to interpret brain signals.
So, communication issues are not yet the most important focus. 
However, we made a simplified survey of technologies that support actual applications.
As today requirements are flexible in terms of latency and throughput, any technology is able to support \ac{B2BC}.

Considering future development in \ac{B2BC}, we extrapolated new potential architectures and technologies that probably will support innovative applications based on wireless technologies.
%
%
Most of these requirements are related to \ac{URLL} communication and simultaneously the network considers scenarios with a high user density.
A new waveform called \ac{GFDM} to support physical layer flexibility.
Other important potential technology that will enable \ac{B2BC} is \ac{NOMA}, specially for scenarios with high interference density.

In summary, \ac{B2BC} is a prospective technology that will gain important improvements and applications that potentially will change the way people transmit information between humans and with things.

%
\vspace{-.2cm}
\bibliographystyle{IEEEtran}
\bibliography{main.bib}
\end{document}